\pdfoutput=1
\documentclass[12pt,a4paper]{article}
 
\usepackage{ifthen} 
\newboolean{pdflatex}
\setboolean{pdflatex}{true} 
 
\newboolean{articletitles}
\setboolean{articletitles}{true} 
 
\newboolean{uprightparticles}
\setboolean{uprightparticles}{false} 
 
\newboolean{inbibliography}
\setboolean{inbibliography}{false} 
 
\usepackage{microtype}
\usepackage{lineno}  
\usepackage{xspace} 
 
\usepackage{graphicx}  
\usepackage{color}
\usepackage{colortbl}
 
\usepackage{amsmath} 
\usepackage{amssymb}
\usepackage{amsfonts}
\usepackage{upgreek} 
 
\newcommand*\patchAmsMathEnvironmentForLineno[1]{%
\expandafter\let\csname old#1\expandafter\endcsname\csname #1\endcsname
\expandafter\let\csname oldend#1\expandafter\endcsname\csname
end#1\endcsname
 \renewenvironment{#1}%
   {\linenomath\csname old#1\endcsname}%
   {\csname oldend#1\endcsname\endlinenomath}%
}
\newcommand*\patchBothAmsMathEnvironmentsForLineno[1]{%
  \patchAmsMathEnvironmentForLineno{#1}%
  \patchAmsMathEnvironmentForLineno{#1*}%
}
\AtBeginDocument{%
\patchBothAmsMathEnvironmentsForLineno{equation}%
\patchBothAmsMathEnvironmentsForLineno{align}%
\patchBothAmsMathEnvironmentsForLineno{flalign}%
\patchBothAmsMathEnvironmentsForLineno{alignat}%
\patchBothAmsMathEnvironmentsForLineno{gather}%
\patchBothAmsMathEnvironmentsForLineno{multline}%
}
 
\usepackage{hyperref}    
\usepackage[all]{hypcap} 
 
\def\lhcb {\mbox{LHCb}\xspace}

\ifthenelse{\boolean{uprightparticles}}%
{

 \def\Ppi         {\ensuremath{\uppi}\xspace}

 \def\Ppsi        {\ensuremath{\uppsi}\xspace}

 \def\PDelta      {\ensuremath{\Delta}\xspace}                 
 \def\PXi      {\ensuremath{\Xi}\xspace}                 
 \def\PLambda      {\ensuremath{\Lambda}\xspace}                 
 \def\PSigma      {\ensuremath{\Sigma}\xspace}                 
 \def\POmega      {\ensuremath{\Omega}\xspace}                 
 \def\PUpsilon      {\ensuremath{\Upsilon}\xspace}                 
 
 \def\PB      {\ensuremath{\mathrm{B}}\xspace}                 
                  
 \def\PD      {\ensuremath{\mathrm{D}}\xspace}

 \def\PJ      {\ensuremath{\mathrm{J}}\xspace}                 
 \def\PK      {\ensuremath{\mathrm{K}}\xspace}

 \def\Pb      {\ensuremath{\mathrm{b}}\xspace}                 
 \def\Pc      {\ensuremath{\mathrm{c}}\xspace}

 \def\Pi      {\ensuremath{\mathrm{i}}\xspace}

 \def\Ps      {\ensuremath{\mathrm{s}}\xspace}

}
{

 \def\Ppi         {\ensuremath{\pi}\xspace}

 \def\Ppsi        {\ensuremath{\psi}\xspace}                 
                  
 \mathchardef\PDelta="7101
 \mathchardef\PXi="7104
 \mathchardef\PLambda="7103
 \mathchardef\PSigma="7106
 \mathchardef\POmega="710A
 \mathchardef\PUpsilon="7107
                  
 \def\PB      {\ensuremath{B}\xspace}                 
                  
 \def\PD      {\ensuremath{D}\xspace}

 \def\PJ      {\ensuremath{J}\xspace}                 
 \def\PK      {\ensuremath{K}\xspace}

 \def\Pb      {\ensuremath{b}\xspace}                 
 \def\Pc      {\ensuremath{c}\xspace}

 \def\Pi      {\ensuremath{i}\xspace}

 \def\Ps      {\ensuremath{s}\xspace}

}

\def\squark    {\ensuremath{\Ps}\xspace}

\def\cquark    {\ensuremath{\Pc}\xspace}

\def\bquark    {\ensuremath{\Pb}\xspace}

\def\pion  {\ensuremath{\Ppi}\xspace}

\def\pip   {\ensuremath{\pion^+}\xspace}

\def\kaon  {\ensuremath{\PK}\xspace}
  \def\Kbar  {\kern 0.2em\overline{\kern -0.2em \PK}{}\xspace}

\def\Kp    {\ensuremath{\kaon^+}\xspace}
\def\Km    {\ensuremath{\kaon^-}\xspace}

\def\Kstarzb {\ensuremath{\Kbar^{*0}}\xspace}

  \def\Dbar    {\kern 0.2em\overline{\kern -0.2em \PD}{}\xspace}

\def\B       {\ensuremath{\PB}\xspace}
\def\Bbar    {\ensuremath{\kern 0.18em\overline{\kern -0.18em \PB}{}}\xspace}

\def\Bz      {\ensuremath{\B^0}\xspace}
\def\Bzb     {\ensuremath{\Bbar^0}\xspace}

\def\Bd      {\ensuremath{\B^0}\xspace}
\def\Bs      {\ensuremath{\B^0_\squark}\xspace}
\def\Bsb     {\ensuremath{\Bbar^0_\squark}\xspace}
\def\Bdb     {\ensuremath{\Bbar^0}\xspace}

\def\jpsi     {\ensuremath{{\PJ\mskip -3mu/\mskip -2mu\Ppsi\mskip 2mu}}\xspace}

  \def\Y#1S{\ensuremath{\PUpsilon{(#1S)}}\xspace}

\def\Lz {\ensuremath{\PLambda}\xspace}
\def\Lbar {\ensuremath{\kern 0.1em\overline{\kern -0.1em\PLambda}}\xspace}

\def\Lb      {\ensuremath{\Lz^0_\bquark}\xspace}

\def\to                 {\ensuremath{\rightarrow}\xspace}

\def\AT#1     {\ensuremath{A_{\mathrm{T}}^{#1}}\xspace}           

\def\C#1      {\ensuremath{\mathcal{C}_{#1}}\xspace}                       
\def\Cp#1     {\ensuremath{\mathcal{C}_{#1}^{'}}\xspace}                    
\def\Ceff#1   {\ensuremath{\mathcal{C}_{#1}^{\mathrm{(eff)}}}\xspace}        
\def\Cpeff#1  {\ensuremath{\mathcal{C}_{#1}^{'\mathrm{(eff)}}}\xspace}       
\def\Ope#1    {\ensuremath{\mathcal{O}_{#1}}\xspace}                       
\def\Opep#1   {\ensuremath{\mathcal{O}_{#1}^{'}}\xspace}                    

\newcommand{\tev}{\ifthenelse{\boolean{inbibliography}}{\ensuremath{~T\kern -0.05em eV}\xspace}{\ensuremath{\mathrm{\,Te\kern -0.1em V}}\xspace}}
\newcommand{\gev}{\ensuremath{\mathrm{\,Ge\kern -0.1em V}}\xspace}
\newcommand{\mev}{\ensuremath{\mathrm{\,Me\kern -0.1em V}}\xspace}
\newcommand{\kev}{\ensuremath{\mathrm{\,ke\kern -0.1em V}}\xspace}
\newcommand{\ev}{\ensuremath{\mathrm{\,e\kern -0.1em V}}\xspace}
\newcommand{\gevc}{\ensuremath{{\mathrm{\,Ge\kern -0.1em V\!/}c}}\xspace}
\newcommand{\mevc}{\ensuremath{{\mathrm{\,Me\kern -0.1em V\!/}c}}\xspace}
\newcommand{\gevcc}{\ensuremath{{\mathrm{\,Ge\kern -0.1em V\!/}c^2}}\xspace}
\newcommand{\gevgevcccc}{\ensuremath{{\mathrm{\,Ge\kern -0.1em V^2\!/}c^4}}\xspace}
\newcommand{\mevcc}{\ensuremath{{\mathrm{\,Me\kern -0.1em V\!/}c^2}}\xspace}

\def\mum  {\ensuremath{{\,\upmu\rm m}}\xspace}

\def\invfb   {\ensuremath{\mbox{\,fb}^{-1}}\xspace}

\def\gsim{{~\raise.15em\hbox{$>$}\kern-.85em
          \lower.35em\hbox{$\sim$}~}\xspace}
\def\lsim{{~\raise.15em\hbox{$<$}\kern-.85em
          \lower.35em\hbox{$\sim$}~}\xspace}

\def\pt         {\mbox{$p_{\rm T}$}\xspace}

\def\tell1  {TELL1\xspace}
\def\ukl1   {UKL1\xspace}

\usepackage{cite} 
\usepackage{mciteplus}
\begin{document}




\begin{titlepage}

\vspace*{-1.5cm}
\centerline{\large EUROPEAN ORGANIZATION FOR NUCLEAR RESEARCH (CERN)}
\vspace*{1.5cm}
\hspace*{-5mm}\begin{tabular*}{16cm}{lc@{\extracolsep{\fill}}r}
\vspace*{-12mm}\mbox{\!\!\!\includegraphics[width=.12\textwidth]{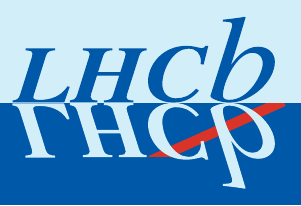}}& & \\ 
& & CERN-PH-EP-2014-027\\
& & LHCb-PAPER-2014-003 \\  
& & \today \\ 
\end{tabular*}

\vspace*{4.0cm}

{\bf\boldmath\huge
\begin{center}
Precision measurement of the ratio of the  $\Lb$ to $\Bzb$ lifetimes
\end{center}
}

\vspace*{2.0cm}

\begin{center}
The LHCb collaboration\footnote{Authors are listed on the following pages.}
\end{center}

\vspace{\fill}

\begin{abstract}
 \noindent
 The LHCb measurement of the lifetime ratio of the \Lb baryon to the \Bzb meson is updated using 
 data corresponding to an integrated luminosity of 3.0~fb$^{-1}$  collected using 7 and 8 TeV centre-of-mass energy $pp$ collisions at the LHC.  The  decay modes used are $\Lb\to\jpsi p K^-$ and 
 $\Bzb\to\jpsi \pi^+ K^-$, where the $\pi^+K^-$ mass is consistent with that of the $\Kstarzb(892)$ meson. The lifetime ratio is determined with unprecedented precision to be $0.974\pm0.006\pm0.004$, where the first uncertainty is statistical and the second systematic. This result is in agreement with original theoretical predictions based on the heavy quark expansion. Using the current world average of the \Bzb lifetime, the \Lb lifetime is found to be $1.479 \pm 0.009 \pm 0.010$~ps.
\end{abstract}

\vspace*{2.0cm}

\begin{center}
  Submitted to Physics~Letters~B 
\end{center}

\vspace{\fill}

{\footnotesize 
\centerline{\copyright~CERN on behalf of the \lhcb collaboration, license \href{http://creativecommons.org/licenses/by/3.0/}{CC-BY-3.0}.}}
\vspace*{2mm}

\end{titlepage}

\pagestyle{empty}  



\setcounter{page}{2}
\mbox{~}
\centerline{\large\bf LHCb collaboration}
\begin{flushleft}
\small
R.~Aaij$^{41}$, 
B.~Adeva$^{37}$, 
M.~Adinolfi$^{46}$, 
A.~Affolder$^{52}$, 
Z.~Ajaltouni$^{5}$, 
J.~Albrecht$^{9}$, 
F.~Alessio$^{38}$, 
M.~Alexander$^{51}$, 
S.~Ali$^{41}$, 
G.~Alkhazov$^{30}$
% ...

{\footnotesize \it
$ ^{1}$Centro Brasileiro de Pesquisas F\'{i}sicas (CBPF), Rio de Janeiro, Brazil\\
% ...
$ ^{5}$Clermont Universit\'{e}, Universit\'{e} Blaise Pascal, CNRS/IN2P3, LPC, Clermont-Ferrand, France\\
% ...
$ ^{9}$Fakult\"{a}t Physik, Technische Universit\"{a}t Dortmund, Dortmund, Germany\\
% ...
$ ^{30}$Petersburg Nuclear Physics Institute (PNPI), Gatchina, Russia\\
% ...
$ ^{37}$Universidad de Santiago de Compostela, Santiago de Compostela, Spain\\
$ ^{38}$European Organization for Nuclear Research (CERN), Geneva, Switzerland\\
% ...
$ ^{41}$Nikhef National Institute for Subatomic Physics, Amsterdam, The Netherlands\\
% ...
$ ^{46}$H.H. Wills Physics Laboratory, University of Bristol, Bristol, United Kingdom\\
% ...
$ ^{51}$School of Physics and Astronomy, University of Glasgow, Glasgow, United Kingdom\\
$ ^{52}$Oliver Lodge Laboratory, University of Liverpool, Liverpool, United Kingdom\\
% ...
}
\end{flushleft}

\cleardoublepage



\pagestyle{plain} 
\setcounter{page}{1}
\pagenumbering{arabic}


%
\clearpage
\section{Introduction}

The heavy quark expansion (HQE) is a
powerful theoretical technique in the description of decays of hadrons containing heavy quarks. 
This model describes inclusive decays and has been used extensively in the analysis of beauty and charm  hadron decays, for example  in the extraction of Cabibbo-Kobayashi-Maskawa matrix elements, such as $|V_{cb}|$ and $|V_{ub}|$ \cite{PDG}.
The basics of the theory were derived in the late 1980's \cite{Shifman:1986mx,*Shifman:1984wx,*Shifman:1986sm,*Guberina:1986gd}. For $b$-flavoured hadrons, the expansion of the total decay width in terms of powers of $1/m_b$, where $m_b$ is the $b$ quark mass, was derived a few years later \cite{Blok:1992hw,*Blok:1992he,*Bigi:1991ir,*Bigi:1992su}. These developments are summarized in Ref.~\cite{Bigi:1994wa}. It was found that there were no terms of  $\mathcal{O}(1/m_b)$, that the  $\mathcal{O}(1/m_b^2)$ terms were tiny, and initial estimates of ${\cal{O}}(1/m^3_b)$ \cite{Neubert:1996we,Uraltsev:1996ta,*DiPierro:1999tb}  effects were small. Thus differences of only a few percent were expected between the \Lb and \Bzb total decay widths, and hence their lifetimes \cite{Neubert:1996we,Cheng:1997xba,Rosner:1996fy}.

In the early part of the past decade, measurements of the ratio of \Lb to \Bzb lifetimes, $\tau_{\Lb}/\tau_{\Bdb}$, gave results considerably smaller than this expectation. In 2003 one experimental average
gave $0.798\pm0.052$~\cite{Battaglia:2003in}, while another was  $0.786\pm0.034$~\cite{Tarantino:2003qw,*Franco:2002fc}. 
Some authors sought to explain the small value of the ratio by including additional operators or other modifications \cite{Ito:1997qq,*Gabbiani:2003pq,*Gabbiani:2004tp,*Altarelli:1996gt}, while others thought that the
HQE could be pushed to provide a ratio of  about 0.9 \cite{Uraltsev:2000qw}, but not so low as the measured value.  
Recent measurements have obtained higher values \cite{Aad:2012bpa,*Chatrchyan:2013sxa,*Aaltonen:2010pj}. In fact, the most precise previous measurement from LHCb, $0.976\pm0.012\pm0.006$ \cite{Aaij:2013oha}, based on 1.0~\invfb of data, agreed with the early HQE expectations.

In this paper we present an updated result for  $\tau_{\Lb}/\tau_{\Bdb}$ using data from 3.0~\invfb of integrated luminosity collected with the LHCb detector from $pp$ collisions at the LHC. Here we add the 2.0~\invfb data sample from the 8~TeV data to our previous 1.0~\invfb 7~TeV sample \cite{Aaij:2013oha}. The data are combined and analyzed together.  Larger simulation samples are used than in our previous publication, and uncertainties are significantly reduced. 

The $\Lb$ baryon is detected in the $\jpsi p\Km$ decay mode, discovered by LHCb \cite{Aaij:2013oha}, while the $\Bdb$ meson is reconstructed in $\jpsi \Kstarzb(892)$ decays, with $\Kstarzb(892)\to\pip\Km$.\footnote{Charge-conjugate modes are implicitly included throughout this Letter.} These modes have the same topology into four charged tracks, thus facilitating cancellation of systematic uncertainties in the lifetime ratio. 

The \lhcb detector~\cite{Alves:2008zz} is a single-arm forward
spectrometer covering the \mbox{pseudorapidity} range $2<\eta <5$,
designed for the study of particles containing \bquark or \cquark
quarks. The detector includes a high-precision tracking system
consisting of a silicon-strip vertex detector surrounding the $pp$
interaction region, a large-area silicon-strip detector located
upstream of a dipole magnet with a bending power of about
$4{\rm\,Tm}$, and three stations of silicon-strip detectors and straw
drift tubes~\cite{Arink:2013twa} placed downstream.
The combined tracking system provides a momentum measurement with
relative uncertainty that varies from 0.4\% at 5\gev to 0.6\% at 100\gev,
and impact parameter resolution of 20\mum for
tracks with large transverse momentum, \pt.\footnote{We use natural units with $\hbar=c=1$.} 
 Different types of charged hadrons are distinguished using information
from two ring-imaging  Cherenkov (RICH) detectors~\cite{LHCb-DP-2012-003}. Photon, electron and
hadron candidates are identified by a calorimeter system consisting of
scintillating-pad and preshower detectors, an electromagnetic
calorimeter and a hadronic calorimeter. Muons are identified by a
system composed of alternating layers of iron and multiwire
proportional chambers~\cite{LHCb-DP-2012-002}.
The trigger~\cite{Aaij:2012me} consists of a
hardware stage, based on information from the calorimeter and muon
systems, followed by a software stage, which applies a full event
reconstruction.

\section{Event selection and $b$ hadron reconstruction}
\label{sec:selection}
Events selected for this analysis are triggered by a $\jpsi\to\mu^+\mu^-$ decay, where the $\jpsi$ is required at the software level to be consistent with coming from the decay of a $b$ hadron by use of either impact parameter (IP) requirements or detachment of the reconstructed $\jpsi$ decay position from the associated primary vertex.

Events are required to pass a cut-based preselection and then are further filtered using a multivariate discriminator based on the boosted decision tree (BDT) technique~\cite{Breiman,*AdaBoost}.  To satisfy the preselection requirements the muon candidates must have \pt larger than 550~MeV, while the  hadron candidates are required to have \pt larger than 250~MeV.  
Each muon is required to have  $\chi^2_{\rm IP}>4$,  where $\chi^2_{\rm IP}$ is defined as the difference in $\chi^2$ of the primary vertex reconstructed with and without the considered track.
Events must have a $\mu^+\mu^-$ pair that forms a common vertex with $\chi^2 < 16$ and that has an invariant mass between $-48$ and +43 MeV of the known \jpsi mass \cite{PDG}. Candidate $\mu^+\mu^-$ pairs are then constrained to the \jpsi mass to improve the determination of the \jpsi momentum. The two charged final state hadrons must have a vector summed \pt of more than 1~GeV, and  form a vertex with $\chi^2/{\rm ndf}<10$, where ndf is the number of  degrees of freedom, and a common vertex with the \jpsi candidate with $\chi^2/{\rm ndf}<16$.   Particle identification requirements are different for the two modes. Using information from the RICH detectors, a likelihood is formed for each hadron hypothesis. The difference in the logarithms of the likelihoods, DLL$(h_1-h_2)$, is used to distinguish between the two hypotheses, $h_1$ and $h_2$ \cite{LHCb-DP-2012-003}. In the \Lb decay the kaon candidate
must have DLL$(K-\pi)>4$ and DLL$(K-p)>-3$, while the proton candidate must have DLL$(p-\pi)>10$ and DLL$(p-K)>-3$.
For the \Bzb decay, the requirements on the pion candidate are DLL$(\pi-\mu)>-10$ and DLL$(\pi-K)>-10$, while DLL$(K-\pi)>0$ is required for the kaon. 

\begin{figure}[!t]
\centering
\includegraphics[width=0.8\textwidth]{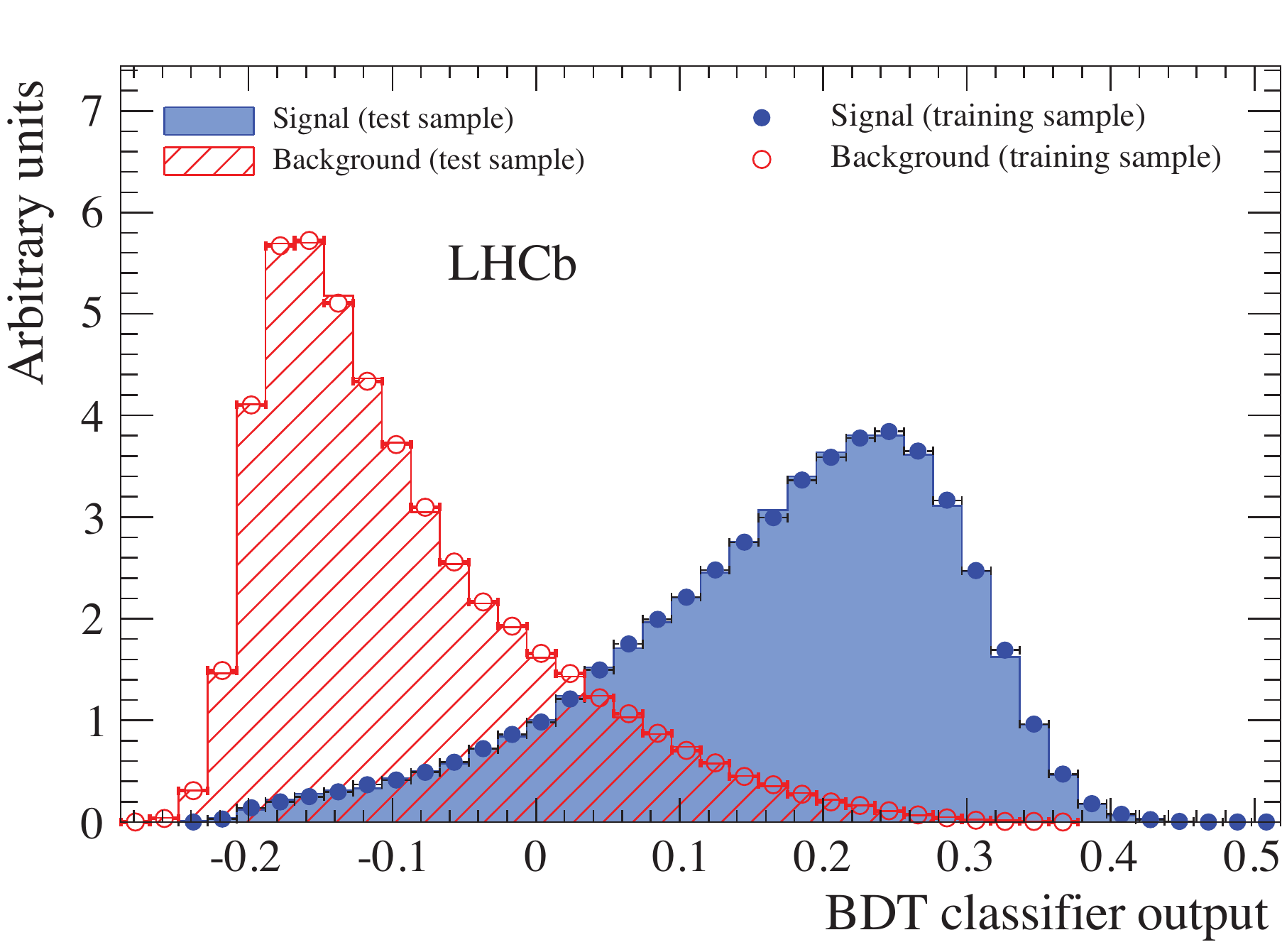}
\caption{\small BDT classifier output for the signal and background. Both training and test samples are shown; their definitions are given in the text.}
\label{fig:BDT_response}
\end{figure}

The BDT selection uses the smaller value of the  DLL($\mu-\pi$) of the $\mu^+$ and $\mu^-$ candidates, the $\pt$ of each of the two charged hadrons,  and their sum,  the \Lb \pt, the \Lb vertex $\chi^2$,  and the $\chi^2_{\rm IP}$ of the \Lb candidate with respect to the primary vertex.
The choice of these variables is motivated by minimizing the dependence of the selection efficiency on decay time; for example, we do not use the $\chi^2_{\rm IP}$ of the proton, the kaon, the flight distance, or the pointing angle of $\Lb$ to the primary vertex.  
To train and test the BDT we use a simulated sample of  $\Lb\to\jpsi p K^-$events for signal  and a background data sample from the mass sidebands in the region $100-200$~MeV below the \Lb  signal peak. Half of these events are used for training, while the other half are used for testing.
The BDT selection is chosen to maximize $S^2/(S+B)$, where  $S$ and $B$ are the signal and background yields, respectively. This optimization includes the requirement that the \Lb candidate decay time be greater than 0.4~ps. The same BDT selection is used for $\Bzb\to\jpsi \pi^-K^+$ decays. The distributions of the BDT classifier output for signal and background are shown in Fig.~\ref{fig:BDT_response}.   The final selection requires that the BDT output variable be greater than 0.04.

\begin{figure}[!t]
\begin{center}
    \includegraphics[width=0.9\textwidth]{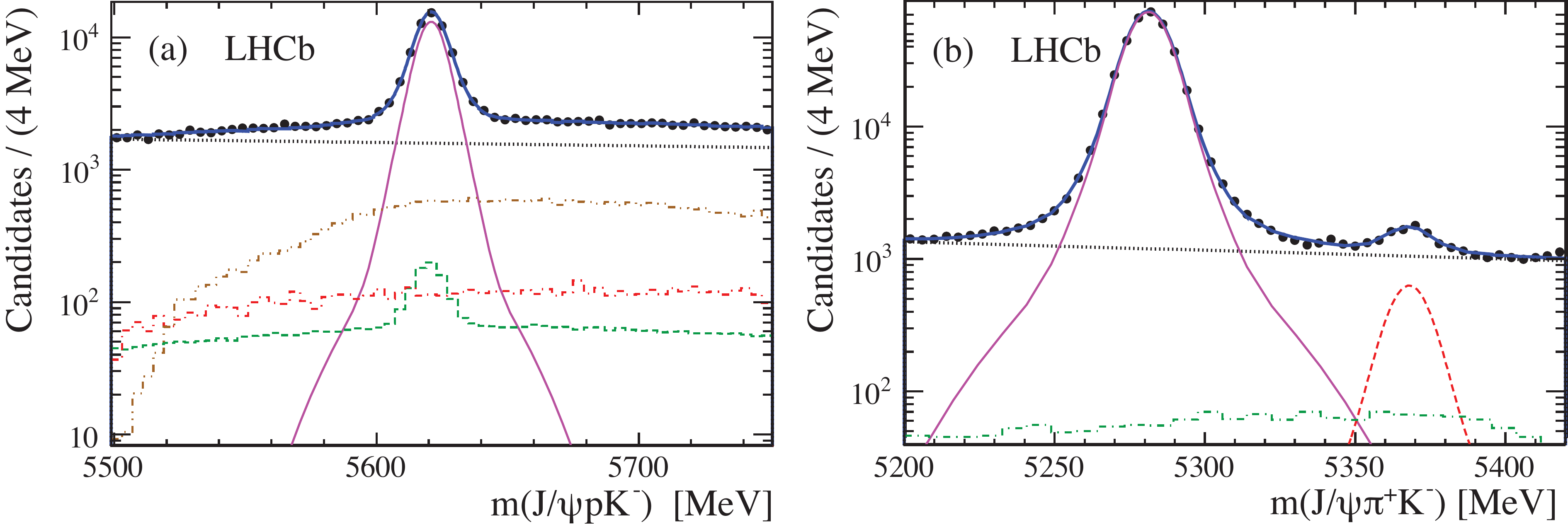}%
\end{center}
\label{fig:signal-log}
\vskip -0.5cm
\caption{\small Fits to the invariant mass spectrum of  (a) $\jpsi pK^-$ and  (b)  $\jpsi \pi^+K^-$ combinations. The $\Lb$ and \Bzb signals are shown by the (magenta) solid  curves. The (black) dotted lines are the combinatorial backgrounds, and the (blue) solid curves show the totals. In (a) the $\Bsb \to \jpsi  K^+K^-$ and $\Bdb \to \jpsi \pi^+K^-$ reflections, caused by particle misidentification, are shown with the (brown) dot-dot-dashed and (red) dot-dashed  shapes, respectively, and the (green) dashed shape represents the  doubly misidentified $\jpsi K^{+}\overline{p}$ final state, where the kaon and proton masses are swapped.  In (b) the $\Bs\to \jpsi \pi^+K^-$ mode is shown by the (red) dashed  curve and the (green) dot-dashed shape represents the $\Lb\to\jpsi p\Km$ reflection.}
\end{figure}

The resulting $\Lb$ and \Bzb candidate invariant mass distributions are shown in Fig.~\ref{fig:signal-log}.  For \Bzb candidates we also require that the invariant $\pi^+ K^-$ mass be within $\pm 100$ MeV of the $\Kstarzb(892)$ mass. 
In order to measure the number of signal events we need to ascertain the backgrounds. The background is dominated by random track combinations at masses around the signal peaks, and their shape  is assumed to be exponential in invariant mass. 
 Specific backgrounds arising from incorrect particle identification, called ``reflections,"  are also considered. In the case of the \Lb decay, these are 
 $\Bsb \to \jpsi K^+K^-$ decays where a kaon is misidentified  as a proton and $\Bdb \to \jpsi\Kstarzb(892)$ decays with $\Kstarzb(892)\to\pip\Km$ where the pion is misidentified as a proton. There is also a double misidentification background caused by swapping the kaon and proton identifications. 
 
\begin{figure}[!b]
\begin{center}
    \includegraphics[width=0.9\textwidth]{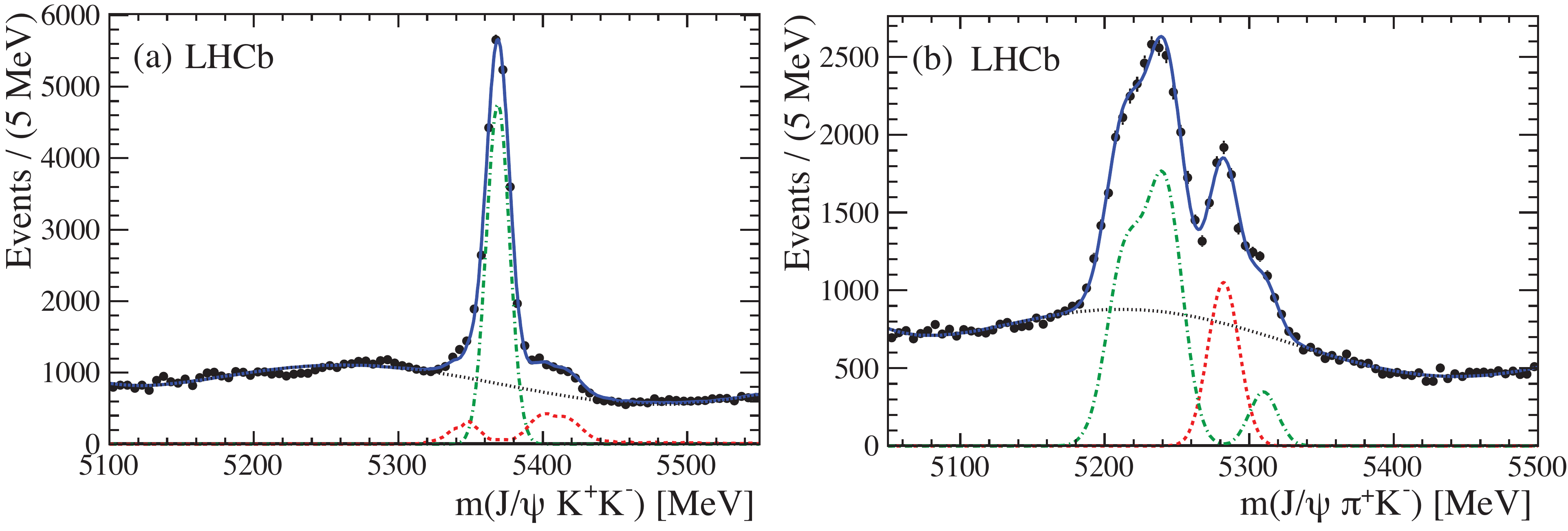}%
\end{center}\label{fig:reflection}
\vskip -0.5cm
\caption{\small Invariant mass distributions of $\jpsi pK^-$ data candidates in the sideband regions $60-200$ MeV on either side of the $\Lb$ mass peak, reinterpreted as misidentified (a) $\Bsb \to \jpsi K^+K^-$  and (b) $\Bdb\to \jpsi \pi^+\Km$ combinations through appropriate mass reassignments. The (red) dashed curves show the $\Bdb$ contributions and  the (green) dot-dashed curves show $\Bsb$ contributions. The (black) dotted curves represent the polynomial background and the (blue) solid curves the total.}
\end{figure}

 To study these backgrounds, we examine the mass combinations in the sideband regions from $60-200$ MeV on either side of the $\Lb$ mass peak.   
 Specifically for each  candidate in the $\jpsi pK^-$   sideband regions we reassign to the proton track the kaon or pion mass hypothesis
 respectively, and plot them separately. The resulting distributions are shown in Fig.~\ref{fig:reflection}.
The $m(\jpsi  K^+K^-)$ invariant mass distribution shows a large peak at the $\Bsb$ mass. There is also a small contribution from the $\Bdb$ final state where the $\pi^+$ is misidentified as a $p$.  The $m(\jpsi\pi^+\Km)$ distribution, on the other hand, shows a peak at the \Bdb mass with  a large contribution from \Bsb decays where the $\Kp$ is misidentified as a $p$.  For both distributions the shapes of the different contributions are determined using simulation. Fitting both distributions we find  19\,327$\pm$309 \Bsb, and 5613$\pm$285 \Bdb events in the \Lb sideband.
 
Samples of simulated  $\Bsb\to\jpsi  K^+K^-$ and $\Bdb\to \jpsi K^-\pi^+$ events are used to find the shapes of these reflected backgrounds in the $\jpsi p K^-$ mass spectrum. Using the event yields found in data and the simulation shapes, we estimate $5603 \pm 90$  $\Bsb\to\jpsi  K^+K^-$ and  $1150 \pm 59$ $\Bzb\to\jpsi \pi^+K^-$ reflection candidates within $\pm 20 $ MeV of the $\Lb$ peak. These numbers are used as Gaussian constraints in the mass fit described below with the central values as the Gaussian means and the uncertainties as the widths.
Following a similar procedure we find $1138\pm 48$ doubly-misidentified  $\Lb$ decays under the $\Lb$ peak.  This number is also used as a Gaussian constraint in the mass fit.

To determine the number of $\Lb$ signal candidates we perform an unbinned maximum likelihood fit to the candidate $\jpsi pK^-$ invariant mass spectrum shown in Fig.~\ref{fig:signal-log}(a). 
The fit function is the sum of the $\Lb$ signal component, combinatorial background, the contributions from the $\Bsb\to\jpsi  K^+K^-$ and $\Bdb \to \jpsi \pi^+K^-$ reflections and the  doubly-misidentified $\overline{\Lb}\to\jpsi K^{+}\overline{p}$ decays. The signal is modeled by a triple-Gaussian function with common means. The fraction and the width ratio  for the  second and third Gaussians  are fixed to the values obtained in the fit to $\Bdb\to\jpsi\Kstarzb(892)$ decays, shown in Fig.~\ref{fig:signal-log}(b). The effective r.m.s.$\!$ width is 4.7 MeV. The combinatorial background is described by an exponential function. 
The shapes of reflections and doubly-misidentified contributions are described by histograms imported from the simulations.  The mass fit gives $50\,233\pm 331$ signal and $15\,842\pm 104$ combinatorial background candidates, $5642\pm 88$  $\Bsb\to\jpsi  K^+K^-$ and $1167\pm 58$ $\Bdb \to \jpsi \pi^+K^-$ reflection candidates, and $1140\pm48$ doubly-misidentified $\Lb$ candidates within $\pm 20$ MeV of the $\Lb$ mass peak. 
The $pK^-$ mass spectrum is consistent with that found previously \cite{Aaij:2013oha}, with a distinct peak
near 1520 MeV, together with the other broad resonant and non-resonant structures that cover the entire kinematic region. 

The $\Bdb$  candidate mass distribution can  be polluted by the reflection from  $\Lb\to\jpsi p\Km$ and $\Bsb\to\jpsi K^+K^-$ decays. 
Following a similar procedure as for the analysis of the \Lb mass spectra,
we take into account the reflection under the $\Bdb$ peak. Figure~\ref{fig:signal-log}(b)  shows the fit to the $\jpsi \pi^+K^-$ mass distribution. There are signal peaks at both $\Bdb$ and $\Bsb$ masses on top of the background. A triple-Gaussian function with common means is used to fit each signal.  The shape of the  $\Bs \to \jpsi \pi^+K^-$ mass distribution is taken to be the same as that of the signal $\Bdb$ decay. The effective r.m.s.$\!$ width is 6.5 MeV. An exponential function is used to fit the combinatorial background. 
The shape of the $\Lb\to\jpsi p\Km$ reflection is taken from simulation, the yield  being Gaussian constrained in the global fit to the expected value.
The mass fit gives $340\,256\pm 893$ signal and $11\,978\pm 153$ background candidates along with a negligible $573\pm27$ contribution of $\Lb\to\jpsi p\Km$ reflection candidates within $\pm 20 $ MeV of the $\Bdb$ mass peak. All other reflection contributions are found to be negligible.


\section{\boldmath Measurement of the $\Lb$ to \Bzb lifetime ratio}
The decay time, $t$, is calculated  as 
\begin{equation}
t = m  \frac{\vec{d} \cdot \vec{p}}{|\vec{p}|^2},
\end{equation}
where $m$ is the reconstructed invariant mass,  $\vec{p}$ the momentum and $\vec{d}$ the flight distance vector of the particle between the production and decay vertices. The  $b$ hadron is constrained to come from the primary vertex.
To avoid systematic biases due to shifts in the measured decay time, we do not constrain the two muons to the \jpsi mass.

The decay time distribution of  the $\Lb\to \jpsi pK^-$ signal can be described by an exponential function convolved with a resolution function, $G(t-t',\sigma_{\Lb})$,  where $t'$ is the true decay time, multiplied by an acceptance function, $A_{\Lb}(t)$:
\begin{equation}
F_{\Lb}(t)=A_{\Lb}(t)\times[e^{-t'/\tau_{\Lb}}\otimes G(t-t',{\sigma_{\Lb}})].
\end{equation}
 The ratio of the decay time distributions of  $\Lb\to \jpsi pK^-$  and $\Bdb \to \jpsi \Kstarzb(892)$ is given by
\begin{equation}
R(t)=\frac{A_{\Lb}(t)\times[e^{-t'/\tau_{\Lb}}\otimes G(t-t',{\sigma_{\Lb}})]}{A_{\Bdb}(t)\times[e^{-t'/\tau_{\Bdb}}\otimes G(t-t',{\sigma_{\Bdb}})]}.\label{eq:ratio}
\end{equation}
The advantage of measuring the lifetime through the ratio is that the decay time acceptances introduced by the trigger requirements, selection and reconstruction almost cancel in the ratio of the decay time  distributions.  The decay time resolutions are 40~fs for the \Lb decay and 37~fs for the \Bzb decay \cite{Aaij:2013oha}.  They are both small enough in absolute scale, and similar enough for differences in resolutions between the two modes not to affect the final result. Thus,
\begin{equation}
R(t)=R(0)e^{-t(1/\tau_{\Lb}-1/\tau_{\Bdb})}=R(0)e^{-t\Delta_{\Lz B}},\label{eq:timefit}
\end{equation}
where $\Delta_{\Lz B}\equiv 1/\tau_{\Lb}-1/\tau_{\Bdb}$ is the width difference and $R(0)$ is the normalization. Since the acceptances are not quite equal, a correction is implemented to first order by modifying Eq.~(\ref{eq:timefit}) with a linear function
\begin{equation}
R(t)=R(0)[1+a t]e^{-t\Delta_{\Lz B}},\label{eq:timefit2}
\end{equation}
where $a$ represents the slope of the acceptance ratio as a function of decay time.


The decay time acceptance is the ratio between the reconstructed decay time distribution for selected events and the generated decay time distribution convolved with the triple-Gaussian decay time resolutions obtained from the  simulations. In order to ensure that the $p$ and $p_{\rm T}$ distributions of the generated $b$ hadrons are correct, we weight the simulated samples to match the data distributions. The simulations do not model the hadron identification efficiencies with sufficient accuracy for our purposes. Therefore we further weight the samples according to the hadron identification efficiencies  obtained from  $D^{*+}\to \pi^+D^0,$ $D^0\to K^-\pi^+$ events for pions and kaons, and $\Lz \to p\pi^-$ for protons.
\begin{figure}[t]
\begin{center}
    \includegraphics[width=0.8\textwidth]{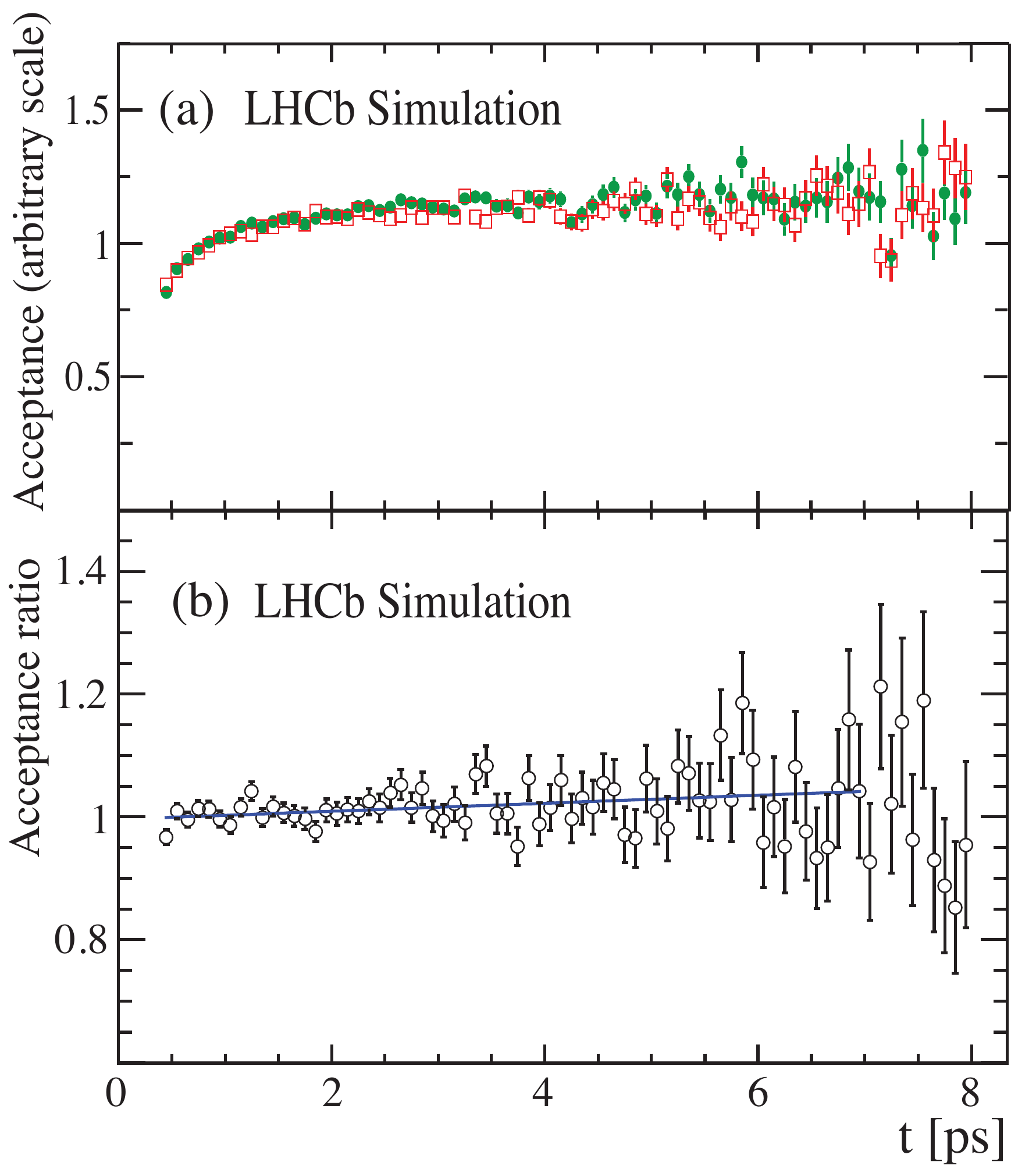}
\end{center}\label{fig:acceptances}
\vskip -0.5cm
\caption{\small (a) Decay time acceptances (arbitrary scale) from simulation for (green) circles $\Lb \to \jpsi pK^-$, and  (red) open-boxes $\Bdb \to \jpsi \Kstarzb(892)$  decays. (b) Ratio of  the decay time acceptances between $\Lb \to \jpsi pK^-$ and  $\Bdb \to \jpsi \Kstarzb(892)$ decays obtained from simulation. The (blue) line shows the result of the linear fit.}
\end{figure}
The $\Lb\to \jpsi pK^{-}$ sample is also weighted using signal yields in bins of $m\left(pK^{-}\right)$.

 The decay time acceptances obtained from the  weighted simulations are shown in Fig.~\ref{fig:acceptances}(a).  The individual acceptances in both cases exhibit the same behaviour.  
The ratio of the decay time acceptances is shown in Fig.~\ref{fig:acceptances}(b). For decay times greater than 7~ps, the acceptance is poorly determined, while below 0.4~ps the individual acceptances decrease quickly. Thus, we consider decay times in the range $0.4-7.0$~ps.
A $\chi^2$ fit to the acceptance ratio with a function of the form $C(1+at)$ between 0.4 and 7~ps, gives a slope $a=0.0066\pm0.0023~\rm ps^{-1}$ and an intercept of $C=0.996\pm0.005$.  The $\chi^2/\rm ndf$ of the fit is $65/64$.  


In order to determine the ratio of  \Lb to \Bzb lifetimes, we determine the yield of $b$ hadrons for both decay modes using unbinned maximum likelihood fits described in Sec.~\ref{sec:selection} to the $b$ hadron mass distributions in 22 bins of decay time of equal width between 0.4 and 7 ps. We use the parameters found from the time integrated fits fixed in each time bin, with the signal and background yields allowed to vary, except for the double misidentification background fraction that is fixed.

\begin{figure}[t]
\begin{center}
    \includegraphics[width=0.9\textwidth]{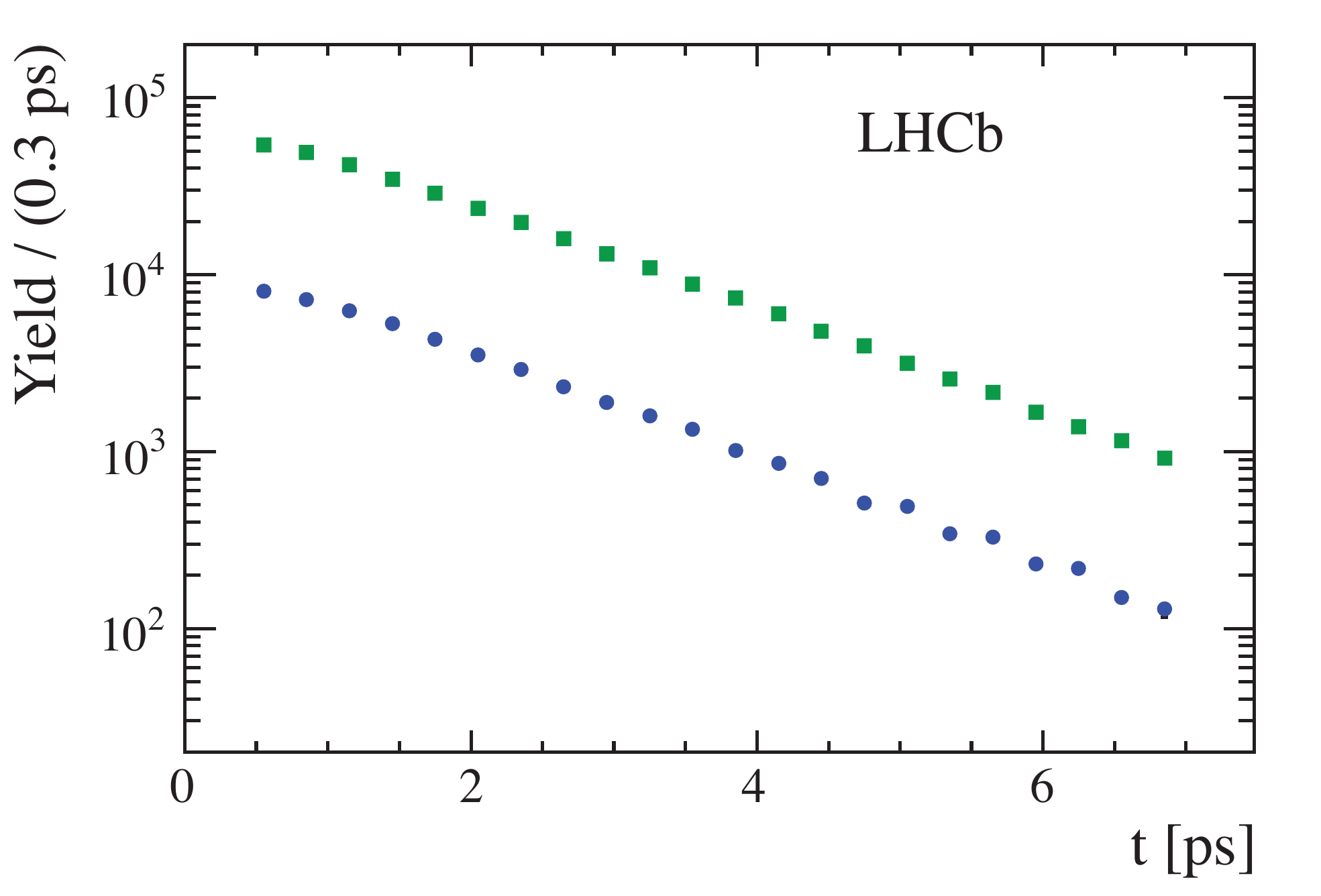}
\end{center}\label{fig:lifetime}
\vskip -0.5cm
\caption{\small Decay time distributions for  $\Lb \to \jpsi pK^-$ shown as (blue) circles, and   $\Bdb \to \jpsi \Kstarzb(892)$ shown as (green) squares. For most entries  the error bars are smaller than the points.}
\end{figure}

\begin{figure}[htb]
\begin{center}
    \includegraphics[width=0.8\textwidth]{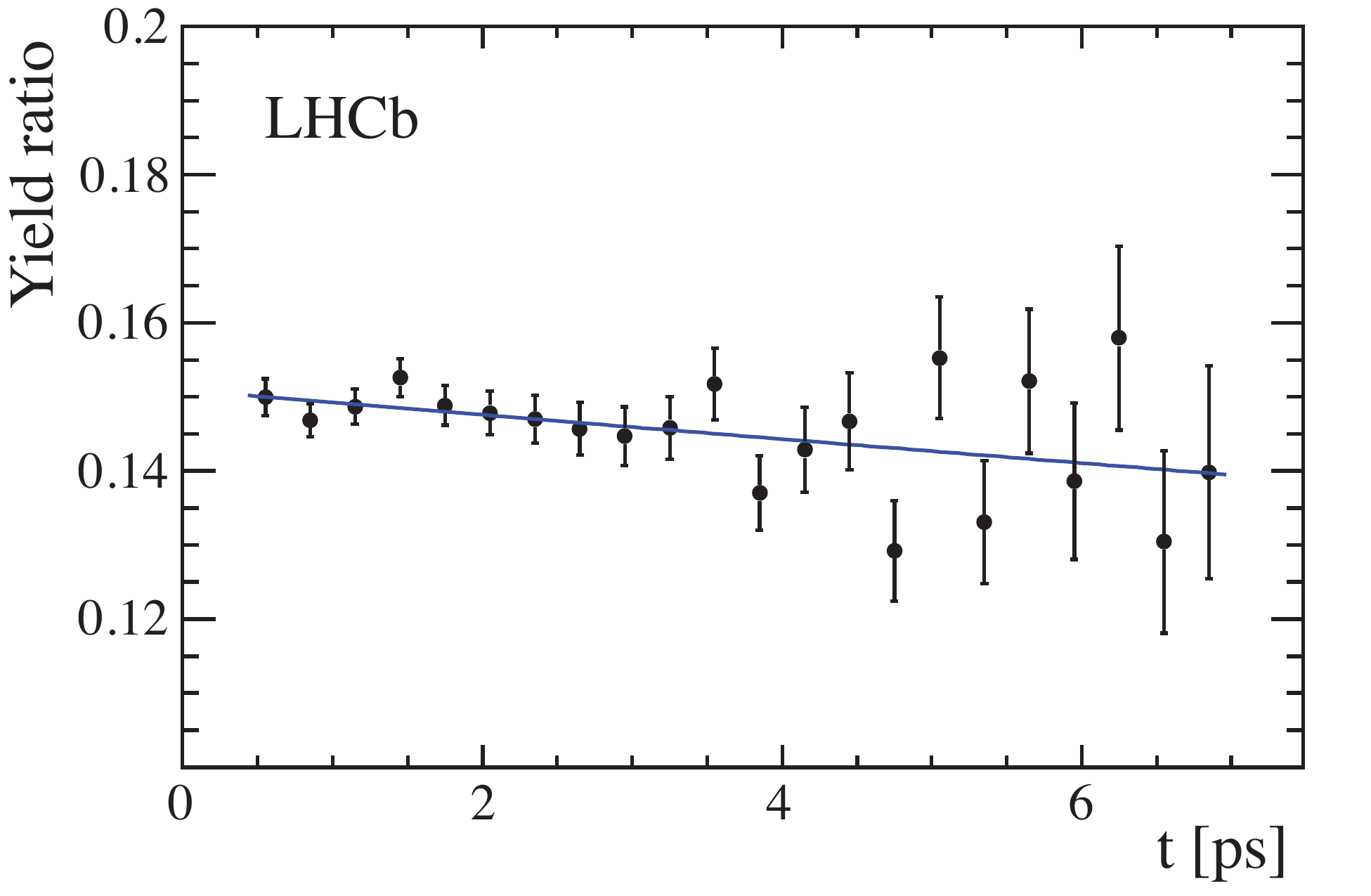}%
\end{center}\label{fig:yield_ratio}
\vskip -0.5cm
\caption{\small Decay time ratio between $\Lb \to \jpsi pK^-$ and  $\Bdb \to \jpsi \Kstarzb(892)$ decays, and the fit for $\Delta_{\Lz B}$ used to measure the $\Lb$ lifetime.}
\end{figure}

The resulting signal yields as a function of decay time are shown in Fig.~\ref{fig:lifetime}.
The subsequent decay time ratio distribution fitted with the function given in Eq.~\ref{eq:timefit2} is shown in Fig.~\ref{fig:yield_ratio}.
A $\chi^2$ fit is used with the slope $a=0.0066~\rm ps^{-1}$ fixed, and both the normalization parameter $R(0)$, and $\Delta_{\Lz B}$ allowed to vary. The fitted value of the reciprocal lifetime difference is
\begin{equation*}
\Delta_{\Lz B} = 17.9 \pm 4.3 \pm 3.1~ \rm ns^{-1}.
\end{equation*}
Whenever two uncertainties are quoted, the first is statistical and second systematic. The latter will be discussed in Sec.~\ref{sec:sys}. The $\chi^2/\rm ndf$ of the fit is $20.3/20$. The resulting ratio of lifetimes is
\begin{equation}
\frac{\tau_{\Lb}}{\tau_{\Bdb}}= \frac{1}{1+ \tau_{\Bdb}\Delta_{\Lz B}} =0.974\pm0.006\pm0.004, \nonumber
\end{equation}
where we use the world average value $1.519\pm 0.007$ ps  for $ \tau_{\Bdb}$ \cite{PDG}. This result is consistent with and more precise than our previously measured value of $0.976\pm0.012\pm0.006$ \cite{Aaij:2013oha}. 
Multiplying the lifetime ratio by  $\tau_{\Bdb}$, the $\Lb$ baryon lifetime is 
 \begin{equation*}
\tau_{\Lb}= 1.479 \pm 0.009 \pm 0.010~ \rm ps.
\end{equation*}
\section{Systematic uncertainties}
\label{sec:sys}
Sources of the systematic uncertainties on  $\Delta_{\Lz B}$, $\tau_{\Lb}/\tau_{\Bdb}$ and the $\Lb$ lifetime are summarized in Table~\ref{tab:sys}. 
\renewcommand{\arraystretch}{1.4}
\begin{table}[!t]
\centering
\caption{\small 
Systematic uncertainties on the $\Delta_{\Lz B}$, the lifetimes ratio $\tau_{\Lb}/\tau_{\Bdb}$ and the $\Lb$ lifetime. The systematic uncertainty associated with $\Delta_{\Lz B}$ is independent of the $\Bdb$ lifetime.}
\vspace{0.2cm}
\begin{tabular}{lcccc}
\hline
 Source&$\Delta_{\Lz B}$ $(\rm ns^{-1})$& $\tau_{\Lb}/\tau_{\Bdb}$ & $\tau_{\Lb}$ $(\rm ps)$  \\
\hline
Signal shape         &1.5   & 0.0021  & 0.0032\\
Background model     &0.7   & 0.0010  & 0.0015  \\
Double misidentification &1.3 &0.0019 & 0.0029 \\
Acceptance slope     &2.2   & 0.0032  & 0.0049\\
Acceptance function  &0.2   & 0.0003  & 0.0004\\
Decay time fit range &0.3   & 0.0004  & 0.0006 \\
$pK$ helicity        &0.3   & 0.0004  & 0.0006\\
$\Bdb$ lifetime      &-     & 0.0001  & 0.0068 \\
\hline   
Total                &3.1   & 0.0044 & 0.0096\\
\hline     
 \end{tabular}
\label{tab:sys}
\end{table}
 The systematic uncertainty due to the signal model is estimated by comparing the results between the default fit with a triple-Gaussian function and a fit with a double-Gaussian function. We find  a change of $\Delta_{\Lz B} = 1.5~\rm ns^{-1}$, which we assign as the uncertainty.  Letting the signal shape parameters free in every time bin results in a change of $0.4~\rm ns^{-1}$. 
The larger of these two variations is taken as the systematic uncertainty on the signal shape.

 The uncertainties due to the background are estimated by comparing the default result to that obtained when we allow the exponential background parameter to float in each time bin. We also replace the exponential background function with a linear function; the resulting difference is smaller than the assigned uncertainty due to floating the background shape. The systematic uncertainty due to the normalization of the double misidentification background is evaluated by allowing the fraction to change in each time bin. 

The systematic uncertainties due to the acceptance slope are estimated by varying the slope, $a$, according to its statistical uncertainty from the simulation. An alternative choice of the acceptance function, where a second-order polynomial is used to parametrize the acceptance ratio between $\Lb\to\jpsi p\Km$ and $\Bdb \to \jpsi\Kstarzb(892)$, results in a smaller uncertainty. 
There is also an uncertainty due to the decay time range used because of the possible change of the acceptance ratio at short decay times. This uncertainty is ascertained by changing the fit range to be $0.7-7.0$ ps and using the difference with the baseline fit. This uncertainty is greatly reduced with respect to our previous publication \cite{Aaij:2013oha} due to the larger fit range, finer decay time bins, and larger signal sample.

In order to correctly model the acceptance, which can depend on the kinematics of the decay, the $\Lb\to\jpsi p\Km$ simulation is weighted according to the $m(pK^-)$ distribution observed in data. As a cross-check, we weight the simulation according to the two-dimensional distribution of  $m(pK^-)$ and $pK^{-}$ helicity angle and assign the difference as a systematic uncertainty. In addition, the PDG value for the $\Bdb$ lifetime, $\tau_{\Bdb}=1.519\pm 0.007~\rm ps$~\cite{PDG}, is used to calculate the $\Lb$ lifetime; the errors contribute to the systematic uncertainty. The total systematic uncertainty is obtained by adding all of the contributions in quadrature.
\section{Conclusions}
\noindent
We determine the ratio of lifetimes of the $\Lb$ baryon and $\Bdb$ meson to be
\begin{equation*}
\frac{\tau_{\Lb}}{\tau_{\Bd}}=0.974\pm0.006\pm0.004.
\end{equation*}
This  is the most precise measurement to date and supersedes our previously published  result~\cite{Aaij:2013oha}. It  demonstrates that the \Lb  lifetime is shorter than the \Bzb lifetime by $-(2.6\pm 0.7)$\%, consistent with the original predictions of the HQE \cite{Shifman:1986mx,Neubert:1996we,Uraltsev:1998bk,Bigi:1995jr,Bigi:1994wa}, thus providing validation for the theory.  
Using the world average measured value for the \Bzb lifetime  \cite{PDG}, we determine
\begin{equation*}
\tau_{\Lb}=1.479\pm 0.009 \pm 0.010 ~\rm ps,
\end{equation*}
which is the most precise measurement to date.

LHCb has also made a  measurement of $\tau_{\Lb}$ using the $\jpsi\Lz$ final state obtaining 
$1.415\pm 0.027\pm 0.006$~ps \cite{Aaij:2014owa}. The two LHCb measurements have systematic uncertainties that are only weakly 
correlated, and we quote an average of the two measurements of
$1.468\pm 0.009\pm 0.008$~ps.


\section*{Acknowledgements}
We are thankful for many useful and
interesting conversations with Prof. Nikolai Uraltsev who contributed greatly to theories describing heavy hadron lifetimes; unfortunately
he passed away before these results were available. We express our gratitude to our colleagues in the CERN
accelerator departments for the excellent performance of the LHC. We
thank the technical and administrative staff at the LHCb
institutes. We acknowledge support from CERN and from the national
agencies: CAPES, CNPq, FAPERJ and FINEP (Brazil); NSFC (China);
CNRS/IN2P3 and Region Auvergne (France); BMBF, DFG, HGF and MPG
(Germany); SFI (Ireland); INFN (Italy); FOM and NWO (The Netherlands);
SCSR (Poland); MEN/IFA (Romania); MinES, Rosatom, RFBR and NRC
``Kurchatov Institute'' (Russia); MinECo, XuntaGal and GENCAT (Spain);
SNSF and SER (Switzerland); NAS Ukraine (Ukraine); STFC (United
Kingdom); NSF (USA). We also acknowledge the support received from the
ERC under FP7. The Tier1 computing centres are supported by IN2P3
(France), KIT and BMBF (Germany), INFN (Italy), NWO and SURF (The
Netherlands), PIC (Spain), GridPP (United Kingdom).
We are indebted to the communities behind the multiple open source software packages we depend on.
We are also thankful for the computing resources and the access to software R\&D tools provided by Yandex LLC (Russia).

\clearpage
\ifx\mcitethebibliography\mciteundefinedmacro
\PackageError{LHCb.bst}{mciteplus.sty has not been loaded}
{This bibstyle requires the use of the mciteplus package.}\fi
\providecommand{\href}[2]{#2}

\end{document}